\documentclass[11pt]{article}

\usepackage[utf8]{inputenc}
\usepackage[T1]{fontenc}
\usepackage[margin=1in]{geometry}
\usepackage{amsmath,amssymb,amsfonts}
\usepackage{mathtools}
\usepackage{booktabs}
\usepackage{tabularx}
\usepackage{array}
\usepackage{enumitem}
\usepackage[expansion=false]{microtype}
\usepackage[hidelinks,breaklinks=true]{hyperref}
\usepackage{url}

\newcommand{\Ascent}{\mathcal{A}}      
\newcommand{\Res}{\mathcal{R}}         
\newcommand{\dd}[2]{\frac{d#1}{d#2}}   

\title{\textbf{Beyond Resilience}\\[0.3em]
\large A Conceptual Framework for Civic Ascent}

\author{
  Alexandros Washburn\\
  \textit{CivicVirtue.ai}
  \and
  Carlo Lipizzi\\
  \textit{Stevens Institute of Technology}
}

\date{Working Paper 2026 (v7, June)}

\begin{document}

\maketitle

\begin{center}
\itshape
``I shall leave my city not less, but greater, better, and more beautiful than it was left to me.''\\
\upshape\normalfont\small
--- The Ephebic Oath of Athens, c.\ 4th century BCE. Preserved in Lycurgus, \textit{Against Leocrates}, 77.
\end{center}

\vspace{1em}

\begin{abstract}
\noindent The resilience literature measures urban performance as recovery: the degree to which a city returns to its pre-shock baseline. This paper develops a stronger concept --- \emph{civic ascent} --- as part of a broader research program on the ethology of coupled agent--environment systems, of which the city is the deepest available empirical instance. Civic ascent is defined as the condition in which a city emerges from shock with higher functional capacity than before. We develop a conceptual framework in the ethological tradition, treating the city as a coupled system of three slow state variables --- \emph{topos} (physical structure), \emph{nomos} (institutional structure), and \emph{hexis} (civic judgment) --- together with a fast affective channel $\delta$ through which shocks to topos and nomos reach hexis. The framework distinguishes three structurally distinct pressures on civic systems: shocks (discontinuities in $T$ or $M$), decay (continuous entropy), and leakage (active extraction of civic surplus into non-civic pools). The ascent condition is that reinforcement from cross-coupling of $T$, $M$, and $H$ exceeds the combined loss from decay and leakage. Post-shock ascent is measured by a normalised improvement index $\Ascent(T)$ applied to a composite civic performance signal $P(t)$ constructed from scale-adjusted key performance indicators, distinguishing intrinsic civic ascent from demographically driven growth. New York City after September 11, 2001, is proposed as the primary empirical case; the operational measurement program is specified in the companion NYC Civic Data Map (Washburn 2026c, 133 KPIs) and executed in Paper 2. The reader for whom only the urban contribution is of interest will find it complete in itself; the reader interested in the larger program will find this paper its formal core.
\end{abstract}

\section{Introduction}

\subsection{The Research Arc}

This paper is part of a research program asking how environments cultivate the agents who inhabit them --- what the program calls the ethology of coupled agent--environment systems. The program treats cities as the deepest available empirical instance of such systems, with ten thousand years of recorded operation, accumulated institutional memory, and physical embodiment of solutions that survived repeated shock. No abstract simulation environment, no game-theoretic sandbox, and no laboratory population match this depth. If the question is how environments shape the agents within them, cities are the environment with the longest depth and the richest empirical inheritance.

Paper 1 establishes the formal apparatus for measuring whether such an environment is ascending --- the conceptual framework presented here. Paper 2 executes the apparatus on the New York City post-9/11 case using the published NYC Civic Data Map. Paper 3 translates the framework into a multi-agent AI evaluation methodology, in which agents are scored not on individual task performance but on their effect on the shared environment over time. The MetaGotham platform, currently hosting agents in development, is the agent-based instantiation through which the layered computational dynamics are made tractable. The reader for whom only the urban contribution is of interest will find it complete in itself; the reader interested in the larger program will find this paper its formal core.

The standard resilience literature asks whether cities recover. This paper asks something harder: whether they ascend.

The question is not rhetorical. Equilibrium in a complex urban system is not a stable destination; it is a transitory state preceding either decline or renewal. Cities that optimize for equilibrium alone lose the capacity to respond generatively to shock --- they survive, perhaps, but they do so by consuming the civic capital that would have enabled ascent. The intuition has been stated informally in practitioner contexts as \emph{sustainability is not sustainable; cities require shocks regularly and hyper-recovery resilience simply to remain viable}.\footnote{Washburn, A. (2025). Sustainability vs.\ Resilience: The Risk Equation for Architects. Lecture delivered at HKS ESG Summit, Dallas, Texas, 30 April 2025.} The present paper proposes the formal framework within which such a claim can be stated precisely and tested empirically. Post-shock performance that exceeds pre-shock baseline is not a statistical outlier or a fortunate accident; it is the signature of a civic system whose reinforcing dynamics exceed its entropic ones, and it can be formally defined, measured, and distinguished from mere demographic buoyancy.

The convergent conclusion --- that superlinear urban dynamics preclude stable equilibrium and require continual renewal --- has been reached from a different methodological direction by Bettencourt et al.\ (2007), whose growth equations imply that cities driven by increasing-returns dynamics ($\beta > 1$) must generate accelerating innovation cycles or collapse. Where Bettencourt et al.\ derive this necessity from power-law scaling and a finite-time singularity in population dynamics, the present framework locates the driver in the ambient frequency and diversity of shocks to which a city is subject --- a coupling that becomes especially relevant given observed rises in shock rates across geopolitical, financial, climatological, and biological domains, and which is examined in Paper 3 through the frequency-domain analysis of civic response. The two frameworks differ in the proposed mechanism of required renewal but share a structural conclusion: resilience, properly understood, is the lower bound of urban performance; its upper bound is ascent.

\subsection{The Ethological Framing}

The apparatus developed here is descriptive before it is predictive. It stands in the tradition of ethological rather than physical modelling --- a tradition in which mathematical structure is used to make the phenomena \emph{legible and comparable} rather than to forecast trajectories. Tinbergen and Lorenz did not begin with predictive dynamical models of animal behavior; they began with careful observation of patterns in context, and produced conceptual scaffolds adequate to the phenomena, with precision accumulating as the field matured. The framework proposed here is analogous in method: a structured descriptive apparatus whose mathematical form is chosen for its capacity to sharpen observation, not to generate forecasts. The reader is invited to approach it in that spirit. Section~\ref{sec:epistemic} states the epistemic status of the model formally; the present note is its intellectual ancestry.

This disposition matters because it changes what the framework is accountable for. A predictive model is accountable for calibration, forecast error, and out-of-sample performance. A descriptive framework in the ethological tradition is accountable for legibility, structural adequacy, and the testability of the observational program it motivates. The contribution of Paper 1 is the apparatus. The empirical program that apparatus motivates is Paper 2. Whether the apparatus ultimately warrants tightening into a predictive model is an open question to be settled by what Paper 2 reveals.

\subsection{Why Cities}

Urban shocks --- earthquakes, floods, economic crises, terrorist attacks --- have long served as natural experiments in civic capacity. What they reveal is not simply whether infrastructure can be rebuilt or population restored, but whether the underlying civic system possesses the adaptive depth to emerge stronger than before. The \emph{bathing-suit problem} (echoing Buffett's observation that only when the tide goes out do you learn who has been swimming without a suit) is well known in practice: many urban weaknesses remain invisible under normal conditions, exposed only when the pool is drained. Shock is the primary instrument of civic revelation.

The existing resilience literature, however, treats the pre-shock baseline as the target (Holling, 1973; Folke, 2006). Recovery to 100 percent is success; anything less is failure; anything more is anomaly. While Grinberger and Felsenstein (2014) recognize that cities may ``bounce forward'' rather than merely bounce back, the literature treats this as one of several possible trajectories rather than the proper criterion of success. This paper proposes that the anomaly is in fact the phenomenon of interest.

The motivation for this paper is partly empirical. The author participated directly in New York City's post-September 11 recovery as a senior official in the city's planning and design apparatus, an experience documented in part in Washburn (2013). What that experience revealed --- the rapid institutional adaptation, the acceleration of civic investment, the measurable improvement in urban indicators across multiple domains --- was not recovery in the narrow sense. It was ascent. The distinction between recovery and ascent was subsequently developed in the author's teaching at Stevens Institute of Technology (2014--2017), where resilience trajectory graphs were used to illustrate the concept to graduate students in systems design and urban planning, during the same period in which the urban resilience literature was independently arriving at similar questions (Grinberger \& Felsenstein, 2014). The question this paper attempts to formalize is whether that observation can be stated with sufficient precision to generalize.

\subsection{Scope of This Paper and Program of Work}

This paper is Paper 1 of a three-paper research program, as set out in Section~1.1.

The present paper establishes the formal apparatus for measuring whether such an environment is \emph{ascending}. A future paper (\emph{The Civic Ascent Test}, Washburn 2026d) translates the framework into a multi-agent AI evaluation methodology, in the agent-based modeling tradition founded by Epstein and Axtell (1996) and extended in Epstein (2006, 2014). It proposes that agents be scored not on individual task performance but on their effect on the shared environment over time --- an inversion of the standard alignment question from ``how do we design agents we can live with'' to ``how do we design environments that reliably cultivate agents we can live with.''

The apparatus developed here is specific to cities in its illustrative content but general in its formal structure: it describes a class of coupled environmental systems in which the ascent of the environment depends on the configuration and engagement of its agents. The portability of the apparatus to other such systems --- including multi-agent computational environments --- is the subject of a future paper and of the broader argument in Washburn (2026a). This paper makes no claim about that portability directly. Its contribution is narrower: to specify the civic ascent apparatus with enough precision that its application in those other domains becomes an open, testable question.

The framework developed here draws on three bodies of work: the dynamical systems tradition in urban complexity science (Bettencourt et al., 2007; Bettencourt, 2013; West, 2017); the resilience literature in urban planning and ecology (Holling, 1973; Folke, 2006; Masten, 2001; Grinberger \& Felsenstein, 2014); and the classical political-philosophical account of civic formation, specifically the Aristotelian triad of place, institution, and character (Washburn, 2013). To these it adds a fourth, less explicit tradition: the ethological method of Tinbergen, Lorenz, and their successors, as applied here to the civic rather than the animal case.

\section{Conceptual Framework}

We model the city as a complex adaptive system governed by three slow state variables together with a fast affective channel. The conceptual triad of physical structure, institutional structure, and civic character as the foundation of urban order draws on a tradition running from Aristotle through contemporary urban design practice (Washburn, 2013).

\begin{table}[h]
\centering
\renewcommand{\arraystretch}{1.3}
\begin{tabularx}{\textwidth}{@{}c l X@{}}
\toprule
\textbf{Variable} & \textbf{Term} & \textbf{Definition} \\
\midrule
$T(t)$ & Topos & The physical and spatial structure of the city --- streets, blocks, buildings, infrastructure, and public space. Topos determines the landscape of physical possibility: where movement, exchange, and encounter are easy or difficult. \\
$M(t)$ & Nomos & The institutional and regulatory structure --- laws, zoning, governance, property rights, and public norms. Nomos determines the landscape of permitted and incentivized action. ($M$ is used rather than $N$ to avoid confusion with population $N$.) \\
$H(t)$ & Hexis & The decision rules, habits, and learned heuristics as modified by civic judgment and applied to the maintenance or reform of Topos and Nomos. \\
$\delta(t)$ & Diathesis & The fast-timescale affective channel through which shocks to $T$ and $M$ reach $H$. $\delta$ represents the emotional disposition of the citizenry at short timescales, driven by discontinuities in $T$ or $M$ and signed by the quality of institutional response. \\
\bottomrule
\end{tabularx}
\caption{The three civic state variables and the fast affective channel.}
\end{table}

Together, topos and nomos generate what we call the \emph{scaffold of choice}: the structured field within which urban actors make daily decisions. The practice of choice within this scaffold, repeated across millions of actors over time, cultivates hexis. Hexis in turn reshapes topos and nomos --- through investment, reform, political action, and design. The city is therefore a self-modifying system:
\[
(T, M) \;\rightarrow\; \text{scaffold of choice} \;\rightarrow\; \text{choices} \;\rightarrow\; H \;\rightarrow\; (T, M)
\]
This is the \emph{civic feedback loop}. Its dynamics determine whether a city ascends, equilibrates, or declines.

\subsection{Hexis as Practical Judgment}

In the language of computation, the civic feedback loop describes the mutual development of hardware (topos), software (nomos), and the cultivated capacity of users (hexis) to operate within a given context and to change it for the better. The metaphor is useful but partial: physical substrate matters in ways the computational framing obscures, and the urban design tradition is right to insist that built form is not arbitrary scaffolding but constitutive of the dynamics it supports. With that caveat, the framing illuminates what hexis is and is not.

Hexis is not the running of a fixed algorithm. It includes decision rules, habits, and learned heuristics, but its defining feature is the meta-capacity to recognize when established patterns should be followed and when they should be departed from --- what Aristotle treats as \emph{phronesis}, the practical judgment component of full ethical disposition, and what is more colloquially termed character. The civic feedback loop is the mechanism by which that capacity is cultivated and reformed across generations.

The distinction is sharpened by recent observations of AI agent behavior, in which agents trained at scale exhibit a meta-capacity to circumvent rules and norms at the agent's discretion (Shapira et al., 2026). Hexis is partly algorithmic --- it includes decision rules, habits, and learned heuristics --- but its defining feature is precisely this meta-capacity, the practical judgment that determines when to follow a rule and when to suspend it. What makes hexis Aristotelian rather than purely computational is the role of the environment. Phronesis cannot be programmed; it can only be cultivated in an environment whose response to behavior functions as a continuous source of feedback and correction. The substrate is the curriculum. Hexis is what grows in agents --- human or artificial --- when they are placed in environments structured well enough to demand and reward practical judgment, and over time modify their behavior in response to the environment's reaction. It is not a procedure that produces good judgment; it is the condition from which good judgment grows.

\subsection{Position Relative to Competing Frameworks}

The civic ascent framework intersects four adjacent literatures. Each addresses a piece of the phenomenon; none of them, taken alone, offers what this framework proposes.

\paragraph{Antifragility.} Taleb's (2012) notion of antifragility names the class of systems that gain from disorder, with cities discussed as species-level exemplars. The concept has been extended to urban planning by Ble\v{c}i\'c and Cecchini (2017, 2020), who distinguish antifragile from ``smart'' cities and enumerate fragilizing factors in planning practice (fragile prediction, excess centralization, over-optimization, over-specialization). Ferrando, Ferrando, and McFarlane (2021) further develop antifragile urbanism as a post-COVID research agenda. Taleb's framework is mechanism-agnostic: it characterizes a response property without proposing a structural account of why some systems possess it. The present framework is complementary --- it attempts to supply the structural account that Ble\v{c}i\'c and Cecchini call for, proposing that the configuration of topos, nomos, and hexis (and the dynamics of decay $L$ and leakage $\Lambda$ among them) determines which cities exhibit antifragile behavior. Civic ascent is related to but distinct from antifragility in three ways: ascent is \emph{directional} (improvement, not merely survival-under-stress); ascent is \emph{structural} (the $T/M/H$ model proposes specific channels of amplification); and ascent is \emph{measurable} via $\Ascent(T)$ on scale-adjusted performance residuals.

\paragraph{The Adaptive Cycle and Panarchy.} Gunderson and Holling (2002) introduced the four-phase adaptive cycle (exploitation $\rightarrow$ conservation $\rightarrow$ release $\rightarrow$ reorganization) as the temporal structure of complex adaptive systems, with panarchy describing cross-scale interactions. Allen et al.\ (2014) and Galaz et al.\ (2022) note that the adaptive cycle remains largely metaphorical --- empirical grounding is limited, and the cycle's predictions about temporal asymmetry have been difficult to test. The civic ascent framework is best understood as operating \emph{within} the reorganization ($\alpha$) phase: civic ascent is what happens when a city passes through the release phase (shock) and reorganizes at a higher level of functional capacity than its prior conservation phase. The adaptive cycle predicts that reorganization \emph{can} produce novel recombinations; the $T/M/H$ framework proposes a structural account of when and why such reorganization exceeds the prior state. The two frameworks are thus complementary: panarchy describes the \emph{temporal structure} of change, while civic ascent describes the \emph{civic conditions} under which reorganization becomes generative. Panarchy also poses a serious challenge to sustained ascent: it predicts that high-potential, high-connectedness states eventually become rigidity traps. Whether sustained civic ascent is possible, or whether it inevitably produces the conditions for its own reversal, is a question this framework leaves open.

\paragraph{Social Capital and Post-Disaster Recovery.} Aldrich (2012), in a comparative study of Tokyo 1923, Kobe 1995, Tamil Nadu 2004, and New Orleans 2005, demonstrates that social capital (bonding, bridging, linking) is more predictive of recovery speed than physical infrastructure or government aid. Aldrich and Meyer (2015) further note that social capital is the only form of capital that can be enhanced during an emergency period. Putnam (2000) provides the foundational social-capital text for American civic life. The present framework treats \emph{hexis} as encompassing (though not reducing to) what this literature calls social capital; the framework's contribution is to place hexis in structural relation to topos and nomos through the civic feedback loop, rather than to treat it as an independent predictor. Aldrich's empirical work provides important potential calibration for the framework, and the four comparative cases constitute a set of candidate additional tests beyond the primary NYC validation case.

\paragraph{Post-Disaster Urban Recovery and ``Bounce Forward.''} Campanella (2006) argues that urban resilience is fundamentally a function of resilient and resourceful citizens, not infrastructure --- a claim that directly anticipates the hexis argument advanced here. Vale and Campanella (2005) document the broad historical pattern that cities often recover from physical destruction; the critical questions concern who returns, how the social fabric reconstitutes, and whether the rebuilt city serves the same population. Grinberger and Felsenstein (2014), using an agent-based simulation of a hypothetical earthquake in Jerusalem, distinguish ``bouncing back'' (to a pre-shock equilibrium) from ``bouncing forward'' (to a new stable configuration) and find that no-policy scenarios reach new equilibria more reliably than full-policy scenarios. Hobor (2015) and Gotham and Campanella (2011) provide panarchy-informed empirical studies of post-Katrina New Orleans. The bounce-forward concept is the closest precedent to civic ascent. The present framework extends it in three ways: it formalizes the distinction between bouncing back and bouncing forward as a specific inequality on $\Ascent(T)$; it proposes a structural account ($T/M/H$) of when each outcome obtains; and its empirical operationalization is directional --- $\Ascent(T)$ measures \emph{how much} a city has advanced or declined, not merely whether it has stabilized in a new configuration.

Taken together: antifragility names the phenomenon; the adaptive cycle describes its temporal structure; social capital identifies a critical driver; the bounce-forward literature provides the closest formal precedent. The framework proposed here attempts to integrate these insights by providing a structural model of the civic conditions under which post-shock reorganization becomes generative rather than merely restorative.

\section{Formal Model}

\subsection{Scope and Epistemic Status of the Model}
\label{sec:epistemic}

Before introducing the formal apparatus, a clarification about what the apparatus is and is not.

The coupled dynamical system introduced in Section~\ref{sec:dynamics} is a \emph{structural hypothesis}, not a predictive model. Its purpose is to specify the \emph{sign and topology} of interactions among state variables --- which variables reinforce which, through which channels, and under what asymmetries between decay and leakage --- not to predict the magnitude of any trajectory. The coupling coefficients ($a_i$, $b_i$, $\gamma_i$) are introduced as general parameters without empirical calibration; the functional forms (particularly the product form $A = T \cdot M \cdot H$ in Section~\ref{sec:ascentfn} and the leakage specifications in Section~\ref{sec:leakage}) are modelling choices, not derived results, and are offered as one member of a family of plausible aggregators (including CES, weighted geometric mean, minimum operator) that share the \emph{weakest-link property} --- that the collapse of any single dimension collapses the whole.

The apparatus is offered in the spirit of Holling's (1973) schematic stability landscapes and adjacent to Epstein's (2006) generative standard for social science, in which the criterion for explanation is the demonstration of mechanisms sufficient to grow the phenomenon under study from plausible local interactions. The present framework does not yet meet that generative standard --- the apparatus is dynamical-systems-flavored rather than agent-based --- but it is intended to be tightened toward it through the agent-based work of MetaGotham (Washburn 2026d and forthcoming). The epistemic work the present apparatus performs is of three kinds. First, it makes the interaction structure precise enough to be criticized. Second, it motivates specific empirical quantities (the composite signal $P(t)$ and the ascent metric $\Ascent(T)$) whose calculation does not depend on knowing the coupling coefficients. Third, it specifies the conditions under which the empirical apparatus of Sections~\ref{sec:signal}--\ref{sec:defining} is a valid estimator of the theoretical quantity $\Ascent$ --- namely, the structural validity of the $T/M/H$ decomposition.

A reader who prefers to treat Section~\ref{sec:dynamics} as a notational convenience rather than a dynamical claim will lose nothing in what follows. The ascent metric $\Ascent(T)$ is well defined as an operational quantity on scale-adjusted KPI residuals, independent of the dynamical motivation.

\subsection{Intrinsic Civic Dynamics}
\label{sec:dynamics}

Let the intrinsic civic state of the city be the vector $X(t) = (T(t), M(t), H(t))$. We model the evolution of the civic system as a coupled dynamical system:
\begin{align}
\dd{T}{t} &= a_1 H + a_2 M - b_1 T - \lambda_T(M,H) \\[4pt]
\dd{M}{t} &= a_3 H + a_4 T - b_2 M - \lambda_M(M,H) \\[4pt]
\dd{H}{t} &= a_5 TM + \delta(t) - b_3 H - \lambda_H(M)
\end{align}
where $a_i > 0$ are reinforcing coupling coefficients, $b_i > 0$ are decay terms representing institutional degradation, physical deterioration, and norm erosion, and $\lambda_i$ are leakage functions representing the diversion of civic energy into non-civic pools. The leakage terms are developed formally in Section~\ref{sec:leakage}. The nonlinear term $TM$ in the hexis equation reflects the scaffold-of-choice mechanism: civic habits are cultivated most powerfully when both physical structure and institutional structure are strong and mutually reinforcing. Consistent with Section~\ref{sec:epistemic}, the signs and topology of these couplings are the load-bearing content of the model; the specific functional forms are modelling choices held constant for tractability.

The term $\delta(t)$ is the fast-timescale affective channel through which shocks to $T$ and $M$ reach civic character. Civic character does not accumulate smoothly as the integral of $TM$; it accumulates episodically, with most consolidation occurring in the weeks and months following shocks, when the affective response of the citizenry is elevated and the conditions of the moment are rapidly written into disposition. $\delta(t)$ captures this fast response. It is driven by discontinuities in $T$ or $M$ and is signed according to the quality of institutional response: effective response ($M$ adapts, $T$ is restored, solidarity is visible) produces positive $\delta$ and hexis gains; institutional failure (incompetence, abandonment, broken trust) produces negative $\delta$ and hexis losses. The asymmetry of $\delta$ in practice --- negative responses to failed institutional response tend to be larger in magnitude and faster in onset than positive responses to successful response --- is consistent with the well-documented negativity bias in human affective and social cognition (Baumeister et al., 2001; Rozin \& Royzman, 2001), and is the mechanism through which hexis is harder to build than to destroy. A civic system can lose in weeks what it built over years.

$\delta$ is empirically approached through observables of civic affective state: polling data on public mood and institutional trust, protest frequency and participation, political participation volatility, media sentiment measures. The composite construction of an observable $\hat{\delta}$ is a task for Paper 2.

Shocks are represented in the dynamical system as discontinuities in $T$ or $M$. A flood reduces $T$ discontinuously; the civic response to that reduction --- the $\delta$ pulse, the subsequent institutional adaptation, the reconstruction trajectory --- unfolds through the coupled dynamics. This separation makes the empirical program more tractable: shocks are measurable as step-changes in state variables, and the civic response to them is measurable as subsequent dynamics.

The system above is the minimum adequate dynamical representation of civic change that this framework requires. Three slow state variables capture the physical, institutional, and dispositional dimensions of civic life. A fast affective channel captures the mechanism by which shocks reach disposition. Two loss channels --- passive decay and active extraction --- capture the two structurally distinct ways civic capacity is diminished. Shocks enter as discontinuities in the state variables themselves. The $a_i$ coefficients encode the coupling structure by which civic experience is converted into civic reinforcement --- the learned weights, in a sense to be developed in later work, of a specific city's civic history. What ascent requires, formally, is that this reinforcement structure produce net positive $dA/dt$ across the post-shock horizon. What ascent requires, substantively, is the subject of the rest of this paper and of the empirical program in Paper 2.

\subsection{The Ascent Function}
\label{sec:ascentfn}

We define the ascent function $A$ as a scalar summary of civic fitness:
\begin{equation}
A = T \cdot M \cdot H
\end{equation}
This specification is one member of a family of aggregators (including CES, weighted geometric mean, minimum operator) that share what may be called the \emph{weakest-link property}: the collapse of any single dimension collapses the whole. The product form is chosen here for analytical convenience in deriving the ascent condition below; the specific choice among members of this family is a modelling decision to be tested empirically and is not a deductive result. Taking the total derivative of $A = T \cdot M \cdot H$ and substituting the state equations of Section~\ref{sec:dynamics} in full yields:
\begin{align}
\dd{A}{t} &= R(T,M,H) - L(T,M,H) + \delta TM - \Lambda(T,M,H) \\[4pt]
R(T,M,H) &= a_1 MH^2 + a_2 M^2 H + a_3 TH^2 + a_4 T^2 H + a_5 T^2 M^2 \\[4pt]
L(T,M,H) &= (b_1 + b_2 + b_3) \cdot TMH
\end{align}
where the leakage field $\Lambda$ is developed in Section~\ref{sec:leakage}. The city ascends when reinforcement and the $\delta$ forcing together dominate decay and leakage; it descends when the reverse holds. The threshold condition $dA/dt = 0$ defines the boundary between civic equilibrium and either ascent or decline.

The $\delta \cdot TM$ term is not incidental. It is the formal account of why institutional quality at the moment of shock is important in the framework. When $\delta > 0$ --- effective institutional response, visible solidarity, restored trust --- the term amplifies ascent at a rate proportional to $T \cdot M$: a city with strong physical and institutional structure gains more from good shock response than a degraded one does. When $\delta < 0$, the same mechanism accelerates decline. The asymmetry noted in Section~\ref{sec:dynamics} --- that negative $\delta$ responses tend to be larger in magnitude and faster in onset than positive ones --- means that civic fitness, once lost through institutional failure at the moment of shock, is not recovered symmetrically. This is the formal basis of the claim that how a city responds to shock matters as much as what it does afterward.

\subsection{Three Pressures on the Civic System}

The civic system is subject to three structurally distinct downward pressures, which the framework distinguishes formally. They differ in timescale, mechanism, and observability, and conflating them --- as the resilience literature frequently does --- obscures the dynamics the framework is designed to reveal.

\paragraph{Shocks} are discontinuities in $T$ or $M$. Hurricane Sandy flooding the subway system, the 2008 financial crisis freezing credit markets, the September 11 attacks destroying physical infrastructure and disrupting institutional routines --- these are all sudden, measurable step-changes in state variables. Shocks are events: locatable in time, quantifiable in magnitude, and revealed in KPI time series as discontinuities rather than as trend changes. A shock can in principle be positive (a discontinuous federal infrastructure investment arriving over months rather than years) but the empirically interesting cases are negative. Shocks are the primary triggers of the dynamics this framework is designed to observe.

\paragraph{Decay} is the continuous downward pressure of entropy on the civic state --- physical deterioration, institutional drift, norm erosion. It operates smoothly, proportional to state, at timescales of years to decades. The decay terms $b_i$ in the state equations of Section~\ref{sec:dynamics} represent this pressure. Decay is the default condition in the absence of maintenance; it does not require agency or malfeasance, only inattention.

\paragraph{Leakage}, developed formally in Section~\ref{sec:leakage}, is the active extraction of civic output into non-civic pools --- patronage hiring, regulatory capture, rent-seeking, the diversion of public investment into personal political advantage. Unlike decay, leakage is state-dependent and requires agency. A civic system can exhibit low decay and high leakage simultaneously --- institutions intact in form but hollowed in function. Unlike shocks, leakage operates while the civic structure appears to function normally; it is not discontinuous but sustained.

The framework treats these three pressures as distinct formal objects. Shocks enter the state equations as discontinuities in $T$ or $M$. Decay enters as the continuous $b_i$ terms. Leakage enters as the state-dependent $\Lambda$ field developed below. Each has its own empirical signature and its own measurement strategy in Paper 2.

\subsection{Leakage: Civic Energy Extraction}
\label{sec:leakage}

As distinguished in Section~3.3, leakage is structurally different from both shocks and decay. The decay terms $b_i$ represent passive entropy --- physical deterioration, institutional drift, norm erosion --- operating continuously in the absence of maintenance. Shocks represent discontinuities in $T$ or $M$ from exogenous events. Leakage represents a third mechanism: the active, state-dependent diversion of civic surplus into non-civic pools --- patronage hiring, regulatory capture, rent-seeking, or the redirection of public investment toward personal political advancement. A city can exhibit low decay and high leakage simultaneously --- its institutions intact in form but hollowed in function. We denote the per-domain leakage rates $\lambda_T, \lambda_M, \lambda_H$ and the aggregate leakage field $\Lambda$.

The classification of specific activities as ``leakage'' rather than as equilibrium outcomes of rational actors reflects a theoretical commitment. Public choice theory (Buchanan \& Tullock, 1962; Tullock, 1967) would model the same phenomena differently --- as predictable outcomes of agents maximising within a given institutional structure, rather than as pathological diversions from a civic ideal. The present framework proceeds from the Aristotelian civic tradition (Washburn, 2013) in which the city has normative purposes --- the cultivation of civic virtue among its inhabitants --- against which extraction is properly diagnosed as diversion. We note this commitment rather than concealing it. Readers who reject the normative frame can reinterpret $\Lambda$ as simply the component of civic output that does not feed back into the $T/M/H$ loop, without loss of mathematical content.

We specify the leakage functions in terms of normalized state variables. Throughout this section $T$, $M$, and $H$ are understood to be normalized on $[0,1]$, so that $\tilde{H} = H$ and $\tilde{M} = M$; the tilde notation is retained where it aids clarity. The leakage rates are then:
\begin{align}
\lambda_T &= \gamma_T \cdot (1 - H) \cdot M \cdot (1 - M) \\[4pt]
\lambda_M &= \gamma_M \cdot (1 - H) \cdot M \cdot (1 - M) \\[4pt]
\lambda_H &= \gamma_H \cdot (1 - M)
\end{align}
where $\gamma_i > 0$ are leakage coefficients reflecting the extractive capacity of non-civic actors in each domain. The aggregate leakage field, consistent with the $\Lambda$ expression in Section~\ref{sec:ascentfn}, is:
\begin{equation}
\Lambda = MH \cdot \lambda_T + TH \cdot \lambda_M + TM \cdot \lambda_H
\end{equation}
which is dimensionally consistent under the normalization declared above.

Three structural properties follow. First, strong hexis suppresses extraction: when $H \to 1$, $\lambda_T$ and $\lambda_M \to 0$. An engaged, watchful citizenry raises the cost of extraction. (This is a structural claim about what the functions yield, not a restatement of the definition of hexis; hexis is defined in Section~2 as a disposition, not as leakage suppression.) Second, the $(1-M) \cdot M$ factor in $\lambda_T$ and $\lambda_M$ captures a non-monotonic relationship between institutional strength and extractive activity: leakage peaks at intermediate institutional strength ($M \approx 0.5$) and vanishes at both extremes. When institutions are fully absent ($M \to 0$) there is nothing to extract; when they are fully intact and functioning ($M \to 1$), the $(1-M)$ term suppresses extraction. The empirically relevant zone is partial capture --- institutions present enough to channel resources but weakened enough to be exploited --- which is precisely the condition the framework is designed to diagnose. Third, leakage generates civic cynicism: $\lambda_H = \gamma_H(1-M)$ captures the withdrawal from participation that follows perceived extraction, and is monotonically increasing as institutional strength declines --- the formal account of civic hollowing. The asymmetry between $\lambda_H$ and the $(\lambda_T, \lambda_M)$ pair is intentional: hexis erodes whenever institutions weaken, regardless of whether active extraction is occurring, because the perception of institutional fragility is itself corrosive to civic disposition.

\subsection{The Ascent Condition}

With leakage included, the rate of civic ascent becomes:
\begin{equation}
\dd{A}{t} = R(T,M,H) - L(T,M,H) - \Lambda(T,M,H)
\end{equation}
where the aggregate leakage field $\Lambda$ is:
\begin{equation}
\Lambda(T,M,H) = MH \cdot \lambda_T + TH \cdot \lambda_M + TM \cdot \lambda_H
\end{equation}
The ascent condition is:
\begin{equation}
\dd{A}{t} > 0 \iff R(T,M,H) > L(T,M,H) + \Lambda(T,M,H)
\end{equation}
Civic ascent requires reinforcement to exceed both natural decay and active extraction. Cities must not only maintain their civic systems against entropy but defend them against capture.

\subsection{Three Routes to Ascent}

The ascent condition $R > L + \Lambda$ reveals three distinct routes by which a city can achieve or recover the ascent regime:

\begin{description}[leftmargin=1.5em,style=nextline]
\item[Raise $R$.] Invest in the quality and character of civic life --- the physical environments, institutional designs, and civic programs that strengthen the coupling of $T$, $M$, and $H$ and amplify their mutual reinforcement.
\item[Reduce $L$.] Maintenance investment --- infrastructure renewal, institutional reform, civic education --- lowers the decay field.
\item[Reduce $\Lambda$.] Suppressing extraction through civic engagement and institutional transparency ensures that civic energy generated by the loop stays in the loop.
\end{description}

Only the first route is generative. The second and third reduce the downward pressures; the first builds the capacity for ascent itself. Maintenance and integrity, however necessary, cannot substitute for the cultivation of civic fitness itself.

\section{Separating Civic Fitness from Demographic Scale}

A central challenge in measuring civic ascent is the confounding of civic fitness with demographic scale. Lagos and Vienna present the problem starkly: Lagos grows rapidly but may not ascend civically; Vienna maintains high civic quality in a demographically contracting national context. Raw population growth is not a reliable indicator of civic virtue.

We treat population $N(t)$ and gross urban scale $S(t)$ as contextual state variables --- influenced by civic fitness but driven also by exogenous demographic forces:
\begin{equation}
\dd{N}{t} = \mu(t) + \nu(t) + \kappa_Q \cdot A - \lambda_C(N, S, T, M)
\end{equation}
where $\mu(t)$ is the macro demographic tide, $\nu(t)$ is regional migration competition, $\kappa_Q$ is a coupling coefficient linking the ascent function $A$ to population attraction, and $\lambda_C(\cdot)$ is a congestion or capacity constraint. To isolate civic performance from demographic context, we construct residuals:
\begin{align}
\rho_N(t) &= N(t) / \hat{N}(t) \\[4pt]
\rho_S(t) &= S(t) / \hat{S}(t)
\end{align}
A city with $\rho_N > 1$ is attracting more people than demographic context predicts; $\rho_N < 1$ indicates underperformance relative to expectation. This separates the city's own civic magnetism from the tide it swims in.

\section{The Observable Civic Performance Signal}
\label{sec:signal}

The ascent function $A$ as defined above is not directly observable. What can be observed are the city's key performance indicators. Following the urban scaling framework of Bettencourt et al.\ (2007, 2013), we model each KPI as a power-law function of population multiplied by a time-varying residual performance signal:
\begin{equation}
K_i(t) = \alpha_i \cdot \hat{N}(t)^{\beta_i} \cdot R_i(t)
\end{equation}
where $\alpha_i$ is a city-specific scaling intercept, $\hat{N}(t)^{\beta_i}$ is the expected value of the KPI given the city's expected population and the scaling exponent $\beta_i$ estimated from cross-sectional urban data, and $R_i(t)$ is the residual performance --- the city's actual output relative to what its demographic opportunity predicts. The use of $\hat{N}(t)$ rather than actual population $N(t)$ is deliberate: it insulates the residual from short-run demographic shocks, ensuring that $R_i(t)$ reflects civic performance rather than transient population displacement. Note that $\alpha_i$ denotes the scaling intercept throughout Sections~5 and~6; it is distinct from the ascent function $A = T \cdot M \cdot H$ defined in Section~\ref{sec:ascentfn}.

The residual terms $R_i(t)$ are closely related to the Scale Adjusted Metropolitan Indicators (SAMIs) of Bettencourt, Lobo, Strumsky, and West (2010). SAMIs measure cross-sectional deviations from expected scaling relationships at a single point in time. The present framework extends the SAMI approach in two ways. First, $R_i(t)$ is a time-varying signal, enabling the tracking of civic performance as a dynamic waveform rather than a static rank. Second, the residuals are proposed to reflect an underlying dynamical structure ($T, M, H$). The SAMI framework of Bettencourt and colleagues provides the measurement baseline --- what performance to expect from population size. The present framework proposes a structural interpretation of the residuals that baseline generates. This interpretation is a hypothesis, not a theorem; Paper 2 begins testing it. The present framework does not replace or improve the Bettencourt scaling analysis; it proposes a structural account of its residuals.

When $R_i(t) > 1$ in a given domain, the city is outperforming what its scale alone would predict. When the composite signal $P(t)$ exceeds its population-predicted baseline $\hat{P}(N(t))$ systematically across multiple domains, the city is generating civic surplus not attributable to demographic buoyancy. This is the observable signature of a city in which structural reinforcement exceeds decay and leakage --- the empirical correlate of $dA/dt > 0$ in the dynamical system of Section~3. The framework does not require a separately named multiplier to interpret this excess; it is read directly from the SAMI residuals.

We construct a composite civic performance signal as a weighted average of residuals:
\begin{equation}
P(t) = \sum_i w_i \cdot R_i(t)
\end{equation}
$P(t)$ is the observable civic waveform --- a time-varying composite of scale-adjusted urban performance. It is not GDP, not population, not any single metric. It is the best available proxy for the latent dynamics of $A(t)$. The weights $w_i$ are a specification choice subject to $\sum w_i = 1$. Paper 2 adopts equal domain weights as its primary specification, with principal-component-derived weights as a robustness check, and reports sensitivity across a reasonable class of weight vectors. The companion NYC Civic Data Map (Washburn 2026c) specifies 133 KPIs across all model terms with source agencies, temporal coverage, and availability codes; Section~7.2 below maps the primary KPIs to components of $P(t)$.

\section{Defining Civic Ascent}
\label{sec:defining}

\subsection{The Ascent Metric}

Let a shock occur at time $t_s$. Define the pre-shock baseline as $P^- = \bar{P}(t_s^-)$, where the overbar denotes a trailing three-year average of $P(t)$ prior to the shock (to reduce sensitivity to business-cycle position and short-run volatility). Let post-shock performance at horizon $T$ be $P^+(T) = P(t_s + T)$. The traditional resilience ratio is:
\begin{equation}
\Res(T) = P^+(T) / P^-
\end{equation}
This yields $\Res = 1$ for full recovery, $\Res < 1$ for incomplete recovery, and $\Res > 1$ for post-shock improvement. The resilience literature treats the last case as an anomaly. We propose it is the central phenomenon. We define the civic ascent metric as:
\begin{equation}
\Ascent(T) = \frac{P(t_s + T) - \bar{P}(t_s^-)}{\bar{P}(t_s^-)}
\end{equation}

\begin{table}[h]
\centering
\renewcommand{\arraystretch}{1.3}
\begin{tabularx}{\textwidth}{@{}c X@{}}
\toprule
$\Ascent(T)$ & \textbf{Civic Interpretation} \\
\midrule
$< 0$ & Net civic decline: the city is weaker after the shock than before. \\
$= 0$ & Resilience: the city has returned exactly to its prior condition. This is the target of the existing literature --- and the lower bound of the present framework. \\
$> 0$ & Civic ascent: the city has emerged stronger than before. This is the central phenomenon of interest. \\
\bottomrule
\end{tabularx}
\caption{Interpretation of the civic ascent metric $\Ascent(T)$. Resilience ($\Ascent = 0$) is the lower bound; ascent ($\Ascent > 0$) is the proper criterion of civic success.}
\end{table}

The use of a trailing three-year average for the pre-shock baseline, rather than the instantaneous pre-shock value, addresses sensitivity to business-cycle position --- a city assessed at the peak of a boom would otherwise register spurious decline against any reasonable counterfactual, while a city assessed in a trough would register spurious ascent through mean reversion alone. Paper 2 further detrends $P(t)$ for secular national trends prior to constructing $\Ascent(T)$.

\subsection{Internal and External Ascent}

Because $P(t)$ reflects both civic fitness and demographic context, we distinguish two forms of ascent. Internal civic ascent measures improvement in intrinsic civic quality:
\begin{equation}
\Ascent_A(T) = \frac{A(t_s + T) - \bar{A}(t_s^-)}{\bar{A}(t_s^-)}
\end{equation}
Demographic ascent measures whether the city improved its competitive position relative to demographic expectation:
\begin{equation}
\Ascent_N(T) = \frac{\rho_N(t_s + T) - \bar{\rho}_N(t_s^-)}{\bar{\rho}_N(t_s^-)}
\end{equation}
A city may exhibit positive internal ascent with negative demographic ascent --- well-governed in an unfavorable context --- or positive demographic ascent with negative internal ascent --- buoyed by a regional boom despite civic weakness. The framework distinguishes these cases, which raw growth metrics obscure.

\subsection{The Central Proposition}

A city exhibits civic ascent if $\Ascent(T) > 0$ --- that is, if post-shock composite civic performance, after controlling for expected population trajectory and urban scaling relationships, exceeds its pre-shock baseline. This condition constitutes empirical evidence for the theoretical condition $dA/dt > 0$, not its direct observation: $P(t)$ is a weighted composite of observable proxies for $A$, not $A$ itself, and the link between the two is a structural hypothesis whose testing is the task of Paper 2. What the metric provides is a measurement discipline --- a refusal to count recovery as success --- grounded in the claim that a city whose reinforcing civic dynamics genuinely exceed its entropic ones will, over time, produce a post-shock performance signal that exceeds its pre-shock level. Whether New York after September 11 satisfies that condition is an empirical question, not a foregone conclusion.

\section{Empirical Program (Paper 2): New York After September 11}

\subsection{The Validation Case}

The empirical execution of the framework is the subject of Paper 2 (\emph{Measuring Civic Ascent: Scale-Adjusted Performance Residuals in New York City, 1996--2011}, Washburn 2026, forthcoming). The present paper specifies the empirical program without executing it; the companion NYC Civic Data Map (Washburn 2026c, 133 KPIs, available at \url{https://doi.org/10.5281/zenodo.19076634}) provides the empirical infrastructure.

New York City's response to the September 11 attacks offers a candidate case for applying the ascent framework. The shock was severe, sudden, and geographically concentrated. The city's pre-shock baseline is well documented across multiple KPI domains. Post-shock indicators --- economic, demographic, physical, institutional, and social --- are available at high resolution through city agency data, federal records, and longitudinal planning documents including the PlaNYC series. Of the 133 KPIs specified in the data map, 111 (83 percent) are immediately public; the remainder require agency requests. Of the publicly available KPIs, 94 have temporal coverage beginning at or before 2001, providing pre-shock baselines for the 9/11 validation window.

The empirical task of Paper 2 has three components. First, construct the expected population and scale trajectories $\hat{N}(t)$ and $\hat{S}(t)$ using national and regional demographic data for the 2001--2010 period, controlling for the effects of the broader post-9/11 national economic slowdown to isolate New York's city-specific performance. Second, estimate scaling exponents $\beta_i$ for each KPI domain using cross-sectional urban data, following the methodology of Bettencourt et al.\ (2007, 2010). Third, construct the composite signal $P(t)$ and measure $\Ascent(T)$ at horizons $T = 2, 5$, and $10$ years. The hypothesis is that $\Ascent(T) > 0$ at the 10-year horizon, driven by structural improvements in topos, nomos, and hexis simultaneously.

\subsection{KPI Mapping to Model Terms}

The NYC Civic Data Map (Washburn 2026c) provides 133 KPIs across all model terms. The table below maps primary KPIs to model terms and $P(t)$ components with data adequacy assessments for the 9/11 validation case.

\begin{table}[h]
\centering
\footnotesize
\renewcommand{\arraystretch}{1.25}
\begin{tabularx}{\textwidth}{@{}X l l l X@{}}
\toprule
\textbf{KPI} & \textbf{Model Term} & \textbf{Coverage} & \textbf{Source} & \textbf{Role in $P(t)$ / $\Ascent(T)$} \\
\midrule
GDP per capita (NYC MSA) & $P(t)$ & 1970--present & BEA & Primary economic output signal; superlinear $\beta > 1$ \\
Felony crime rate (index) & $P(t)$ / $H$ & 1965--present & NYPD CompStat & Safety output; continued decline post-9/11 is key $\Ascent$ signal \\
Local election turnout & $P(t)$ / $H$ & 1950--present & NYC BOE & Civic participation; Nov 2001 election a hexis event \\
Patent filings per capita & $P(t)$ & 1975--present & USPTO & Innovation output; superlinear $\beta$ \\
Life expectancy at birth & $P(t)$ / $H$ & 1980--present & NYC DOHMH & Wellbeing signal; sublinear $\beta$ \\
4-year graduation rate & $P(t)$ / $H$ & 1990--present & NYC DOE & Human capital formation signal \\
Net domestic migration rate & $\rho_N$ / $\Ascent_N$ & 1990--present & IRS SOI & Demographic ascent: voting with feet \\
Housing Vacancy Survey & $T(t)$ & 1965--present & Census / NYC HPD & Physical stock quality; gold-standard longitudinal \\
Deferred capital maintenance backlog & $T(t)$ / $L$ & 1985--present & NYC OMB & Topos decay measurement; reduced post-9/11 \\
Lower Manhattan residential units & $T(t)$ & 1999--present & NYC DCP / DOB & Topos transformation signal (LMDC conversion) \\
Bond credit rating & $M(t)$ & 1975--present & NYC Comptroller & Institutional confidence; recovery within 18 months \\
Capital budget execution rate & $M(t)$ & 1985--present & NYC OMB & Institutional delivery capacity; Bloomberg surge \\
DOI investigations opened & $M(t)$ / $\Lambda$ & 1980--present & NYC DOI & Leakage suppression signal; oversight activity \\
LMDC commitments vs.\ disbursements & $\Lambda_T$ & 2002--2012 & LMDC Annual Reports & Direct leakage test: procurement gap in capital recovery \\
Sole-source contract share (capital) & $\Lambda_T$ & 2010--present & NYC Comptroller & Procurement leakage indicator; suppression = ascent signal \\
\bottomrule
\end{tabularx}
\caption{Primary KPI mapping to model terms and $P(t)$ components for the NYC 9/11 validation case. Full 133-KPI specification in Washburn (2026c).}
\end{table}

\subsection{$\Ascent(T)$ Sub-Hypotheses at 2, 5, and 10 Years}

The NYC post-9/11 case generates three testable sub-hypotheses at increasing temporal horizons. These are stated here as the target of Paper 2; no empirical results are reported in the present paper.

\paragraph{At $T = 2$ years (2003):} The 2-year horizon tests the hexis and nomos response. Crime rate continued declining ($H$-signal). The November 2001 election turnout spike was a transient hexis event indexed to a $\delta$ pulse following the shock. Bond rating recovered by 2002 ($M$-signal). The hypothesis is $\Ascent(2) \approx 0$ or slightly negative --- too short for structural signals to accumulate, but hexis and nomos responses visible.

\paragraph{At $T = 5$ years (2006):} The 5-year horizon captures early Bloomberg-era capital investment and the LMDC reconstruction programme. Lower Manhattan residential conversion underway. Capital budget execution elevated. The hypothesis is $\Ascent(5) > 0$, driven primarily by nomos and topos signals.

\paragraph{At $T = 10$ years (2011):} The 10-year horizon is the primary test. Full PlaNYC infrastructure cycle. Sustained crime decline. GDP and patent residuals elevated vs.\ peer metros. The hypothesis is $\Ascent(10) > 0$ across all three state variable domains simultaneously --- the signature of full civic ascent rather than single-domain recovery.

\section{Discussion}

\subsection{Growth is Not Virtue}

A city whose population grows because of a regional boom, or because of national demographic pressure, has not necessarily improved its civic fitness. The residualization procedure separates what the city has done from the tide it swims in.

\subsection{Shrinkage is Not Failure}

A city in a contracting demographic context --- a shrinking post-industrial American city, a Japanese regional center --- can exhibit positive internal ascent ($\Ascent_A > 0$) even as its population declines. The framework provides a basis for evaluating such cities on their own terms.

\subsection{Resilience is the Floor, Not the Ceiling}

The traditional target of recovery to baseline is a minimum condition, not a measure of success. Planning, investment, and governance frameworks that define success as $\Res = 1$ are setting an unnecessarily low standard. There is a further danger in baseline-as-target: over successive shocks, a system that accepts recovery-to-prior-baseline as the definition of success progressively redefines what counts as acceptable. Each successful recovery lowers the implicit reference point for the next. The mechanism is now identifiable in the framework: each shock produces a $\delta$ response, and when the institutional response is perceived as merely restorative rather than constructive, the $\delta$ response is neutral or mildly negative, and hexis accumulates cynicism rather than solidarity. A city that has absorbed three shocks and returned to the level it occupied before each has not maintained its civic capacity; it has spent three $\delta$ events on recovery signaling rather than on ascent formation, and it has done so while reporting success. The ascent criterion $\Ascent(T) > 0$ is in part an epistemic corrective --- a measurement discipline that refuses the redefinition of success as mere non-failure. In developmental psychology, Masten (2001) identifies \emph{thriving} --- positive adaptation exceeding prior functioning --- as the upper bound of resilience in human development; the urban analogue proposed here is civic ascent. The Ephebic standard --- leaving the city greater, better, and more beautiful than it was received --- expresses as civic commitment what this framework attempts to express as measurable criterion.

\subsection{Shock as Measurement Instrument}

Many dimensions of civic fitness are invisible under normal conditions. Institutional redundancy, social trust, adaptive governance capacity --- these are latent until stress makes them manifest. Shock is therefore not merely a disturbance to be minimized; it is the primary instrument by which civic depth becomes observable.

A further conjecture regarding \emph{civic universality classes} --- whether cities may occupy distinct persistent ranges of civic coupling strength with systematically different SAMI patterns --- is deferred to future empirical work.

\subsection{The Equity Constraint}

The civic ascent metric $\Ascent(T) > 0$ is a necessary but not sufficient condition for genuine civic ascent. A composite signal $P(t)$ can rise --- and the formal ascent condition $R > L + \Lambda$ can be satisfied in aggregate --- while the distribution of civic gains narrows. This is not merely an ethical concern; it is a structural one. The leakage suppression mechanism identified in Section~\ref{sec:leakage} depends on strong hexis and hexis depends on broad civic participation. A city in which GDP residuals are rising while inequality widens, racial wealth gaps persist and poverty concentrates is a city whose aggregate $\tilde{H}$ may be elevated while its distribution is hollowing. Extraction becomes easier, not harder, when civic engagement is concentrated in a thin stratum. Measured ascent without distributional breadth is structurally fragile: it satisfies the inequality on paper while eroding the conditions that would sustain it.

This implies a distributional adequacy condition as a companion to $\Ascent(T) > 0$: civic ascent is robust only when the residual improvement in $P(t)$ is broad across the income and demographic distribution of the city, not concentrated in its upper quantiles. The NYC Civic Data Map includes income inequality (Gini coefficient), racial wealth gap and poverty rate as inputs to $P(t)$ precisely for this reason, not as normative additions to the framework but as structural diagnostics of whether the hexis underpinning ascent is genuinely broad or spuriously aggregate. Equity-weighted variants of $P(t)$ are a direction for Paper 2 and a necessary condition for the full civic ascent claim.

The NYC post-9/11 case is not exempt from this test. The post-shock period that the framework proposes as a candidate ascent case was also a period of rising real estate values, accelerating gentrification in lower Manhattan and adjacent neighborhoods and documented displacement of lower-income residents. Whether the hexis signal in that period reflects a broadly held civic consolidation or a more concentrated one is an open empirical question, and the answer would qualify or strengthen the $\Ascent(10) > 0$ hypothesis materially.

\section{Conclusion}

This paper has proposed a conceptual framework for civic ascent --- a concept that reframes resilience as the lower bound of urban performance rather than its definition. The framework rests on three elements: a dynamical model of the city as a coupled system of topos, nomos, and hexis with a fast affective channel $\delta$, offered as structural hypothesis rather than predictive model; a scalar ascent function $A = T \cdot M \cdot H$ derived from those variables; and an observable ascent metric $\Ascent(T)$ applied to a composite of scale-adjusted performance signals.

The central theoretical proposition is the ascent condition $R > L + \Lambda$ --- civic ascent requires that the reinforcement field generated by cross-coupling of the state variables exceed both natural decay and active leakage. The asymmetry between decay (passive, slow to respond to investment) and leakage (active, state-dependent, suppressed by strong hexis) is the formal basis of the civic feedback loop: investment in civic life raises $H$, suppresses the conditions under which leakage is possible, and allows the reinforcement field to compound into ascent. The three pressures distinguished in Section~3.3 --- shocks as discontinuities, decay as continuous entropy, leakage as active extraction --- together constitute the full downward pressure on the civic system, each with its own mechanism and empirical signature.

The central empirical claim is that the ascent condition is observable through post-shock KPI residuals, and that New York after September 11 is a candidate case of a city satisfying it. The operational execution of that test is the subject of Paper 2 (Washburn 2026, forthcoming), supported by the companion NYC Civic Data Map (Washburn 2026c) comprising 133 KPIs across all model terms with dedicated leakage series.

Cities that persist across centuries are not cities that have achieved balance; they are cities that have repeatedly metabolized shock into renewal. Equilibrium is not the condition of civic health but the signature of civic pause --- a transitory state between one ascent and the next, or between ascent and decline. The measurement apparatus proposed here is designed to distinguish these cases, and to make visible the difference between a city that has recovered and a city that has advanced. The highest form of resilience is not recovery. It is ascent.

\section*{Data Availability}

The NYC Civic Data Map underlying this paper is available at \url{https://doi.org/10.5281/zenodo.19076634}. The dataset comprises 133 KPIs across eight model terms (Topos, Nomos, Hexis, Decay, Leakage, Shock, Demographic, and Index) with temporal coverage, source agencies, and availability codes for each series. The dataset will be updated as the empirical program proceeds.


\section*{Notation Summary}

\begin{table}[h]
\centering
\small
\renewcommand{\arraystretch}{1.25}
\begin{tabularx}{\textwidth}{@{}c X@{}}
\toprule
\textbf{Symbol} & \textbf{Definition} \\
\midrule
$T(t)$ & Topos: physical and spatial structure of the city \\
$M(t)$ & Nomos: institutional and regulatory structure ($M$ used to avoid confusion with population $N$) \\
$H(t)$ & Hexis: civic habits, behavioral dispositions, and character of the population \\
$\delta(t)$ & Fast-timescale affective channel in the hexis equation; driven by discontinuities in $T$ or $M$, signed by the quality of institutional response \\
$X(t)$ & Intrinsic civic state vector: $X(t) = (T(t), M(t), H(t))$ \\
$A$ & Ascent function: $A = T \cdot M \cdot H$ \\
$dA/dt$ & Rate of civic ascent; positive when reinforcement $R$ exceeds decay $L$ and leakage $\Lambda$ \\
$R(T,M,H)$ & Reinforcement field: $a_1 MH^2 + a_2 M^2 H + a_3 TH^2 + a_4 T^2 H + a_5 T^2 M^2$ \\
$L(T,M,H)$ & Decay field: $(b_1 + b_2 + b_3) \cdot TMH$ \\
$\Lambda(T,M,H)$ & Aggregate leakage field: $\Lambda = MH \cdot \lambda_T + TH \cdot \lambda_M + TM \cdot \lambda_H$ \\
$\lambda_T, \lambda_M, \lambda_H$ & Leakage functions in each domain \\
$\gamma_T, \gamma_M, \gamma_H$ & Leakage coefficients: $\gamma_i > 0$ \\
$\tilde{H}, \tilde{M}$ & Normalized hexis and nomos on $[0,1]$; used in leakage specifications \\
$a_i$ & Reinforcing coupling coefficients in the state equations ($a_i > 0$); encode the coupling structure by which civic experience is converted into civic reinforcement \\
$b_i$ & Decay coefficients in the state equations ($b_i > 0$) \\
$N(t)$ & Actual population at time $t$ \\
$\hat{N}(t)$ & Expected population trajectory from national/regional demographic conditions \\
$\rho_N(t)$ & Demographic residual: $\rho_N = N(t) / \hat{N}(t)$ \\
$\rho_S(t)$ & Scale residual: $\rho_S = S(t) / \hat{S}(t)$ \\
$\kappa_Q$ & Coefficient linking ascent function $A$ to population attraction in the demographic equation of Section~4 \\
$K_i(t)$ & The $i$-th key performance indicator at time $t$ \\
$\beta_i$ & Urban scaling exponent for KPI $i$ (Bettencourt / West) \\
$R_i(t)$ & Residual performance signal: $R_i(t) = K_i(t) / [\alpha_i \cdot \hat{N}(t)^{\beta_i}]$ \\
$w_i$ & Weight for KPI $i$ in composite signal; subject to $\sum w_i = 1$ \\
$P(t)$ & Composite civic performance signal: $P(t) = \sum_i w_i R_i(t)$ \\
$\hat{P}(N)$ & Population-predicted baseline of $P(t)$ from cross-sectional urban scaling \\
$\bar{P}(t_s^-)$ & Trailing three-year average of $P(t)$ before the shock, used as pre-shock baseline \\
$t_s$ & Time of shock \\
$\Res(T)$ & Traditional resilience ratio: $\Res = P^+(T) / \bar{P}(t_s^-)$ \\
$\Ascent(T)$ & Civic ascent metric: $\Ascent(T) = [P(t_s + T) - \bar{P}(t_s^-)] / \bar{P}(t_s^-)$ \\
$\Ascent_A(T)$ & Internal civic ascent: improvement in $A$ post-shock \\
$\Ascent_N(T)$ & Demographic ascent: improvement in $\rho_N$ post-shock \\
\bottomrule
\end{tabularx}
\caption{Principal notation. The fast affective channel $\delta(t)$ carries the forcing previously assigned to exogenous terms, and shocks are represented as discontinuities in the state variables themselves rather than as additive forcing terms.}
\end{table}


\begin{thebibliography}{99}
\small

\bibitem{aldrich2012} Aldrich, D. P. (2012). \textit{Building Resilience: Social Capital in Post-Disaster Recovery}. Chicago: University of Chicago Press.

\bibitem{aldrich2015} Aldrich, D. P., \& Meyer, M. A. (2015). Social capital and community resilience. \textit{American Behavioral Scientist}, 59(2), 254--269.

\bibitem{allen2014} Allen, C. R., Angeler, D. G., Garmestani, A. S., Gunderson, L. H., \& Holling, C. S. (2014). Panarchy: Theory and application. \textit{Ecosystems}, 17, 578--589.

\bibitem{baumeister2001} Baumeister, R. F., Bratslavsky, E., Finkenauer, C., \& Vohs, K. D. (2001). Bad is stronger than good. \textit{Review of General Psychology}, 5(4), 323--370.

\bibitem{bettencourt2007} Bettencourt, L. M. A., Lobo, J., Helbing, D., K\"uhnert, C., \& West, G. B. (2007). Growth, innovation, scaling, and the pace of life in cities. \textit{Proceedings of the National Academy of Sciences}, 104(17), 7301--7306.

\bibitem{bettencourt2010} Bettencourt, L. M. A., Lobo, J., Strumsky, D., \& West, G. B. (2010). Urban scaling and its deviations: Revealing the structure of wealth, innovation and crime across cities. \textit{PLOS ONE}, 5(11), e13541.

\bibitem{bettencourtwest2010} Bettencourt, L. M. A., \& West, G. B. (2010). A unified theory of urban living. \textit{Nature}, 467, 912--913.

\bibitem{bettencourt2013} Bettencourt, L. M. A. (2013). The origins of scaling in cities. \textit{Science}, 340(6139), 1438--1441.

\bibitem{blecic2017} Ble\v{c}i\'c, I., \& Cecchini, A. (2017). On the antifragility of cities and of their buildings. \textit{City, Territory and Architecture}, 4(3).

\bibitem{blecic2020} Ble\v{c}i\'c, I., \& Cecchini, A. (2020). Antifragile planning. \textit{Planning Theory}, 19(2), 172--192.

\bibitem{buchanan1962} Buchanan, J. M., \& Tullock, G. (1962). \textit{The Calculus of Consent: Logical Foundations of Constitutional Democracy}. Ann Arbor: University of Michigan Press.

\bibitem{campanella2006} Campanella, T. J. (2006). Urban resilience and the recovery of New Orleans. \textit{Journal of the American Planning Association}, 72(2), 141--146.

\bibitem{epstein2006} Epstein, J. M. (2006). \textit{Generative Social Science: Studies in Agent-Based Computational Modeling}. Princeton, NJ: Princeton University Press.

\bibitem{epstein2014} Epstein, J. M. (2014). \textit{Agent\_Zero: Toward Neurocognitively Plausible Foundations for Generative Social Science}. Princeton, NJ: Princeton University Press.

\bibitem{epstein1996} Epstein, J. M., \& Axtell, R. (1996). \textit{Growing Artificial Societies: Social Science from the Bottom Up}. Cambridge, MA: MIT Press / Brookings Institution Press.

\bibitem{ferrando2021} Ferrando, D. T., Ferrando, T., \& McFarlane, C. (2021). Towards an antifragile urban form: A research agenda for advancing resilience in the built environment. \textit{URBAN DESIGN International}, 26, 135--157.

\bibitem{folke2006} Folke, C. (2006). Resilience: The emergence of a perspective for social-ecological systems analyses. \textit{Global Environmental Change}, 16(3), 253--267.

\bibitem{galaz2022} Galaz, V., Duit, A., Biggs, R., et al.\ (2022). Panarchy: Ripples of a boundary concept. \textit{Ecology \& Society}, 27(3):21.

\bibitem{gotham2011} Gotham, K. F., \& Campanella, R. (2011). Coupled vulnerability and resilience: The dynamics of cross-scale interactions in post-Katrina New Orleans. \textit{Ecology \& Society}, 16(3), 12.

\bibitem{grinberger2014} Grinberger, A. Y., \& Felsenstein, D. (2014). Bouncing back or bouncing forward? Simulating urban resilience. \textit{Proceedings of the Institution of Civil Engineers: Urban Design and Planning}, 167(3), 115--124.

\bibitem{gunderson2002} Gunderson, L. H., \& Holling, C. S., eds.\ (2002). \textit{Panarchy: Understanding Transformations in Human and Natural Systems}. Washington, DC: Island Press.

\bibitem{hobor2015} Hobor, G. (2015). New Orleans' remarkably (un)predictable recovery: Developing a theory of urban resilience. \textit{American Behavioral Scientist}, 59(10), 1214--1230.

\bibitem{holling1973} Holling, C. S. (1973). Resilience and stability of ecological systems. \textit{Annual Review of Ecology and Systematics}, 4, 1--23.

\bibitem{lycurgus} Lycurgus. \textit{Against Leocrates}, 77. In \textit{Minor Attic Orators}, Vol.\ 2, trans.\ J. O. Burtt. Cambridge, MA: Harvard University Press (Loeb Classical Library), 1954.

\bibitem{masten2001} Masten, A. S. (2001). Ordinary magic: Resilience processes in development. \textit{American Psychologist}, 56(3), 227--238.

\bibitem{putnam2000} Putnam, R. D. (2000). \textit{Bowling Alone: The Collapse and Renewal of American Community}. New York: Simon \& Schuster.

\bibitem{rozin2001} Rozin, P., \& Royzman, E. B. (2001). Negativity bias, negativity dominance, and contagion. \textit{Personality and Social Psychology Review}, 5(4), 296--320.

\bibitem{shapira2026} Shapira, N., et al.\ (2026). Agents of Chaos. arXiv:2602.20021.

\bibitem{taleb2012} Taleb, N. N. (2012). \textit{Antifragile: Things that Gain from Disorder}. New York: Random House.

\bibitem{tullock1967} Tullock, G. (1967). The welfare costs of tariffs, monopolies, and theft. \textit{Western Economic Journal}, 5(3), 224--232.

\bibitem{vale2005} Vale, L. J., \& Campanella, T. J., eds.\ (2005). \textit{The Resilient City: How Modern Cities Recover from Disaster}. Oxford: Oxford University Press.

\bibitem{washburn2013} Washburn, A. (2013). \textit{The Nature of Urban Design: A New York Perspective on Resilience}. Washington, DC: Island Press.

\bibitem{washburn2026a} Washburn, A. (2026a). \textit{Chatting with Barbarians: Can Cities Civilize AI?} Manuscript in preparation.

\bibitem{washburn2026b} Washburn, A. (2026b). Civic frequency: A spectral decomposition of urban civic performance signals. CivicVirtue.ai Working Paper (forthcoming).

\bibitem{washburn2026c} Washburn, A. (2026c). NYC civic data map: 133 KPIs across all model terms. CivicVirtue.ai Technical Document. \url{https://doi.org/10.5281/zenodo.19076634}

\bibitem{washburn2026d} Washburn, A. (2026d). The civic ascent test: An environment-centric framework for evaluating alignment in multi-agent AI systems. CivicVirtue.ai Working Paper (forthcoming).

\bibitem{west2017} West, G. B. (2017). \textit{Scale: The Universal Laws of Growth, Innovation, Sustainability, and the Pace of Life in Organisms, Cities, Economies, and Companies}. New York: Penguin Press.

\end{thebibliography}
\end{document}